\documentstyle[12pt,epsf]{article}
\newcommand{\fsigma}{\mbox{$f_{0}$(400--1200)}}
\newcommand{\fn}{\mbox{$f_{0}$(980)}}
\newcommand{\ft}{\mbox{$f_{0}$(1370)}}
\newcommand{\ff}{\mbox{$f_{0}$(1500)}}
\newcommand{\fs}{\mbox{$f_{0}$(1710)}}

\begin{document}\baselineskip .7cm
\title{\bf Identifying the quark content of the isoscalar scalar mesons \fn,
\ft, and \ff\ from weak and electromagnetic processes \\[5mm]}
\author{Frieder Kleefeld$^{\,a}\!\!$
\footnote{{\tt kleefeld@cfif.ist.utl.pt} (corresponding author)}\,,
Eef van Beveren$^{\,b}\!\!$
\footnote{\tt eef@teor.fis.uc.pt}\,,
George Rupp$^{\,a}\!\!$
\footnote{\tt george@ajax.ist.utl.pt}\,, and
Michael D.\ Scadron$^{\,c}\!\!$
\footnote{\tt scadron@physics.arizona.edu} \\[5mm]
$^{a}${\footnotesize\it Centro de F\'{\i}sica das Interac\c{c}\~{o}es
Fundamentais, Instituto Superior T\'{e}cnico, P-1049-001 Lisboa,
Portugal}\\ [.3cm]
$^{b}${\footnotesize\it Departamento de F\'{\i}sica, Universidade de Coimbra,
P-3004-516 Coimbra, Portugal}\\ [.3cm]
$^{c}${\footnotesize\it Physics Department, University of Arizona, Tucson,
AZ 85721, USA} \\ [.3cm]
[.3cm]
{\small hep-ph/0109158}
}
\date{\today}
\maketitle

\begin{abstract}
The assignments of the isoscalar scalar mesons \fn, \ft, and \ff\
in terms of their flavor substructure is still a matter of heated dispute.
Here we employ the weak and electromagnetic decays
$D_s^+\rightarrow f_0\pi^+$ and $f_0\rightarrow \gamma\gamma$, respectively,
to identify the \fn\ and \ff\ as mostly $\bar{s}s$,
and the \ft\ as dominantly $\bar{n}n$, in agreement with previous work.
The two-photon decays can be satisfactorily described with
quark as well as with meson loops, though the latter ones provide a less
model-dependent and more quantitative description.
\end{abstract}
\section{Introduction}
A proper classification of the scalar mesons is still being
clouded by two major problems, which mutually hamper the resolution of either.
The first difficulty is the apparent excess of experimentally confirmed
scalar resonances with respect to the number of theoretically expected
$\bar{q}q$ states.
The second problem is to unambiguously identify the $\bar{q}q$
configuration of the isoscalar scalar mesons, i.e., the \fsigma\
(or $\sigma$), \fn, \ft, \ff, and \fs.
In previous work, especially the former issue has been addressed, showing
that the light (below 1~GeV) scalars can be described as a complete nonet of
$\bar{q}q$ states,
resulting from either the dynamical breaking of chiral symmetry \cite{S82}, or
the coupling of bare $P$-wave $\bar{q}q$ systems to the meson-meson continuum
in a unitarized approach \cite{BRMDRR86,BR01}.
We believe that these two mechanisms are intimately related to one another,
though in a not yet completely understood fashion.
In any case, in both pictures the scalar mesons between 1.3 and
1.5 GeV form another nonet, and so forth. 
So we conclude there is no excess of observed resonances, thus
dispensing with the introduction of new degrees of freedom.

Here, we want to focus on the second issue, namely the
identification of the isoscalars, especially the vehemently disputed \fn, \ft,
and \ff, in an as model-independent way as one may achieve. In
Refs.~\cite{BR99a,RBS01} qualitative arguments from observed hadronic decays
have already been presented that favor, in our view, a mainly $\bar{s}s$
configuration for the \fn\ and \ff, and a dominantly nonstrange $\bar{q}q$
content for the \ft. Furthermore, we are engaged in substantiating these
arguments by
analysing also the four-pion decays of these scalars via intermediate 
$\rho\rho$ and $\sigma\sigma$ two-resonance states, in a similar way as done
for the $\omega\rightarrow\rho\pi\rightarrow\pi\pi\pi$ cascade process in
Ref.~\cite{KBR01}. In the present work, we shall employ the weak and
electromagnetic decays (as opposed to the more complicated strong-interaction
dynamics) $D_s^+\rightarrow f_0\pi^+$
and $f_0\rightarrow \gamma\gamma$, respectively, which will give quantitative
support for our $\bar{q}q$ assignments. These processes will be analysed
in a simple $\bar{q}q$ picture for the corresponding $f_0$ resonances, with
a minimum of model-dependent input.

This paper is organized as follows. In Section~\ref{sec2} we compute
the weak decays $D^+_s\rightarrow \pi^+ f_0(980)$, $\pi^+ f_0(1500)$, 
$\pi^+ f_0(1710)$ using $W^+$ emission. In Section~\ref{sec3} \/we calculate
the $f_0(980)$, $f_0(1370)\rightarrow 2\,\gamma$ electromagnetic
decays, employing quark as well as meson loops. Conclusions are drawn in
Section~\ref{sec4}.  

\section{Weak decays $D^+_s\rightarrow \pi^+ f_0$} \label{sec2}
First we compute the parity-conserving
weak decays $D^+_s\rightarrow\pi^+ f_0(980)$ and $\pi^+ f_0(1500)$, supposing
for the moment that both of 
these final-state scalar mesons are purely $\bar{s}s$. Given the Fermi
Hamiltonian density $H_W = \frac{G_F}{2\sqrt{2}} \; (J\,J^+ + J^+ J)$ with
\cite{PDG2000} $G_F = 1.16639(1) \cdot 10^{-5}\;\;\mbox{GeV}{}^{-2}$ and
$F_\pi = f_{\pi^+}/\sqrt{2} \simeq (92.42 \pm 0.27)$~MeV, the magnitudes of
the corresponding weak decay amplitudes of $W^+$ emission are \cite{PVA}
(also see Ref.~\cite{S84b})
\begin{eqnarray}
|M(D_s^+\rightarrow \pi^+f_0(980))| & = & 
\frac{G_F \, | V_{ud} | \, | V_{cs} |}{2} \; F_\pi \;
(m^2_{D_s^+}-m^2_{f_0(980)}) \nonumber \\
 & & \nonumber \\
 & = & (159 \pm 24) \cdot 10^{-8}\; \mbox{GeV}, \label{xeq1} \\ 
 & & \nonumber \\
|M(D_s^+\rightarrow \pi^+f_0(1500))| & = & 
\frac{G_F \, | V_{ud} | \, | V_{cs} |}{2} \; F_\pi \;
(m^2_{D_s^+}-m^2_{f_0(1500)}) \nonumber \\
 & & \nonumber \\
 & = & (89 \pm 13) \cdot 10^{-8}\; \mbox{GeV}, \label{xeq2}
\end{eqnarray}
being both close to the data \cite{PDG2000} $(178 \pm 40) \cdot 10^{-8}\;
\mbox{GeV}$ and $(96 \pm 28) \cdot 10^{-8}\; \mbox{GeV}$, respectively. The
latter amplitudes are extracted from the observed decay rates $\Gamma$
according to $|M| = m_{D^+_s} \sqrt{8 \pi\, \Gamma /q_{cm}}$. The agreement of
Eqs.~(\ref{xeq1}) and (\ref{xeq2}) with the data, which has already been noted
in Refs.~\cite{RBS01} and \cite{BRS00}, respectively, shows that first-order
perturbative weak graphs have impressive predictive power.

The formulae of Eqs.~(\ref{xeq1}) and (\ref{xeq2}) are based on the standard
description of weak interactions in terms of Fermi theory, which is a
low-energy tree-level approximation of the Standard Model Lagrangian, or in
other words, a lowest-order description in the spirit of Wilson's Operator
Product Expansion (OPE).
In the language of Ref.~\cite{Buch96}, we only consider the current-current
operator $Q_{2}$ multiplied by the Wilson coefficient $C_{2}$.
Higher orders could be included by taking into account further operators,
$Q_{1}$, $Q_{3}$, $Q_{4}$, $Q_{5}$, $Q_{6}$, multiplied by the corresponding
Wilson coefficients $C_{1}$, $C_{3}$, $C_{4}$, $C_{5}$, $C_{6}$.
From Ref.~\cite{Buch96} we learn that the corresponding contributions are
suppressed and often negative for $K$ and $D$ decays.
Throughout this work we assume $C_{2}=1$, thereby absorbing the anticipated
negative marginal contributions of the further operators as a correction
to the value $C_{2}\approx1.25$ quoted in Ref.~\cite{Buch96}.
Furthermore, we may observe that, since decay rates of $\bar{q}q$ systems are
to a good approximation proportional to $\bar{q}q$ probability distributions
at the $\bar{q}q$ center of mass \cite{Roger67}, the higher-order OPE terms
seem to cancel corrections from the $\bar{q}q$ wave function, such that we meet
the experimental data.

The coincidence that both effects --- one perturbative and one nonperturbative
--- compensate each other may have some physical roots.
It is also important to notice that, although we do not rely on
wave functions in this paper, we bear in mind the nonet assignment
given in Refs.~\cite{BRMDRR86,BR01}, which classifies both the light nonet
of scalar resonances and the nonet between 1.3 and 1.5 GeV as ground states,
each from a different origin.
As a consequence, we do not foresee the usual suppression factors for radial
excitations in the case of the \ff\ (and also the \ft), as for instance used in
Ref.~\cite{AAN01}.

Another way to study Eqs.\ (\ref{xeq1}) and (\ref{xeq2}) above is to take the
ratio
\begin{equation} \left| \,\frac{M(D_s^+\rightarrow \pi^+f_0(980))}{M(D_s^+
\rightarrow \pi^+f_0(1500))} \, \right|_{\,|f_0> = |\bar{s}s>} \; = \;
\frac{m^2_{D_s^+}-m^2_{f_0(980)}}{m^2_{D_s^+}-m^2_{f_0(1500)}} \; = \; 1.79
\pm 0.04, \label{xeq3}
\end{equation}
(using $m_{f_0(980)} = (980 \pm 10)$ MeV, $m_{f_0(1500)} = (1500 \pm 10)$ MeV
and $m_{D_s^+} = (1968.6 \pm 0.6)$ MeV), which is  independent of the weak
scale $G_F$, the CKM parameters $| V_{ud} |$, $| V_{cs} |$, and the pion decay
constant $F_\pi$. As such, Eq.\ (\ref{xeq3}) is the kinematic
(model-independent) infinite-momentum-frame (IMF) (see e.g.\ Ref.~\cite{S92})
version. The data
\cite{PDG2000} depend on the
branching ratio and center-of-mass (CM) momenta as
\begin{equation} \left| \,\frac{M(D_s^+\rightarrow \pi^+f_0(980))}{M(D_s^+
\rightarrow \pi^+f_0(1500))} \, \right|_{\scriptsize \mbox{PDG}} \; = \;
\sqrt{\frac{\Gamma (D_s^+\rightarrow \pi^+f_0(980)) \;\; q_{cm} (D_s^+
\rightarrow \pi^+f_0(1500))}{\Gamma (D_s^+\rightarrow \pi^+f_0(1500))\;\,
q_{cm} (D_s^+\rightarrow \pi^+f_0(980))}} \; = \; 1.86 \pm 0.68\; ,\label{xeq4}
\end{equation}
showing again a very good agreement.
Here, we have used the measured branching ratios \cite{PDG2000}
\mbox{$\Gamma (D_s^+\rightarrow \pi^+f_0(980))/\Gamma (D_s^+) =
(1.8 \pm 0.8) \%$} and \mbox{$\Gamma (D_s^+\rightarrow
\pi^+f_0(1500))/\Gamma (D_s^+) = (0.28 \pm 0.16) \%$}, and the corresponding
extracted CM momenta $q_{cm} (D_s^+\rightarrow \pi^+f_0(980)) =
(732.1 \pm 5.1)$ MeV/c and  $q_{cm} (D_s^+\rightarrow \pi^+f_0(1500)) =
(393.8 \pm 8.1)$~MeV/c. The large error $\pm \,0.68$ in Eq.\ (\ref{xeq4}) stems
from the uncertainties in the measured branching ratios, rather than from
the quite accurately known CM momenta. These uncertainties leave quite some
room to allow for significant $\bar{n}n$ admixtures in the \fn\ as well as the
\ff, without calling into question their $\bar{s}s$ dominance. On the other
hand, from the failure to observe the decay \mbox{$D_s^+\rightarrow
\pi^+f_0(1370)$} \cite{PDG2000} (see however Ref.~\cite{G00}) it seems safe to
conclude that the $f_0(1370)$ does not have a large $\bar{s}s$ component. 

To conclude the weak processes, let us look at the situation for the \fs.
Although the weak decay $D^+_s\rightarrow\pi^+ f_0(1710)$ has been observed,
the quoted rate $(1.5 \pm 1.9)\times 10^{-3}$ \cite{PDG2000}, corresponding to
an amplitude of $(97 \pm 123) \cdot 10^{-8}$ GeV, only accounts for $K^+K^-$
decays of this resonance. The theoretical $W^+$-emission amplitude has a
magnitude of $52 \cdot 10^{-8}$ GeV, if we again ignore possible corrections
from the internal $\bar{q}q$ wave function of the \fs, which may be
questionable for this probably excited state. Also in view of the huge
experimental error,
no definite conclusions on the $\bar{q}q$ (or any other) substructure of the
\fs\ are possible for the time being. Nevertheless, the sheer observation of
the weak decay process seems to preclude a dominantly $\bar{n}n$ configuration.
Indeed, the Meson Particle Listings conclude that the \fs\ \em ``is consistent
with a large $\bar{s}s$ component'' \em \/(Ref.~\cite{PDG2000}, page 470).

\section{Electromagnetic scalar decays $S\rightarrow 2\, \gamma$} \label{sec3}
An alternative process to analyse the flavor content of the $f_0$ mesons is
the two-photon decay, since the corresponding amplitude is very sensitive to
the masses and especially the charges of the particles involved. Moreover, this
process may also provide a tool to determine whether some of these isoscalar
scalar meson are in fact glueballs \cite{P00}. In our analysis, we shall
restrict ourselves to those $f_0$ states for which two-photon decays have been
observed.
\subsection{The decay $f_0(980) \rightarrow 2\,\gamma$}
The PDG tables \cite{PDG2000} now report the scalar
$f_0(980)\rightarrow 2\,\gamma$ decay rate as \makebox{$(0.39\, \pm
\,0.12)$~keV}. Given the scalar amplitude structure
\cite{DEMSB94,DLS99,BBIS00,LN99} $\,M \, \varepsilon_{\mu} (k^{\,\prime}) \,
\varepsilon_{\nu} (k) \, (g^{\,\mu\nu} k^{\,\prime}\cdot k - k^{\,\prime \,\mu}
k^{\,\nu})$, the two-photon decay rate is 
\begin{equation}\Gamma (f_0\rightarrow 2\,\gamma) =
\frac{m_{\,f^{\,0}}^{\,3} \, |M|^2}{64\,\pi}\, , \;\; \mbox{or} \quad
|M(f_0(980)\rightarrow 2\,\gamma)| = (0.91\pm0.14)\cdot
10^{-2}\;\mbox{GeV}^{-1} \; .\label{eq1}
\end{equation} 
If the $f_0(980)$ were $\bar{n}n$, the isoscalar $u$,$d$ quark-loop analogue of
the isovector $\pi^{\,0}\rightarrow 2\,\gamma$ amplitude, given by
\cite{DLS99} $\sqrt{2}\;\alpha\,N_c\,$Tr$\,[Q^{2\,} \, Q_{\bar{n}n}] \,
/\,(\pi\,F_\pi) = 5\,\alpha\,N_c\,/\,(9\,\pi\,F_\pi) \simeq 0.042$~GeV${}^{-1}$
with $N_c=3$, would generate an $f_0(980)\rightarrow 2\,\gamma$ decay rate a
factor of \underline{21 times too large} \footnote{We introduced the SU(3)
charge matrix $Q= T_3 + Y/2 =$ Diag$\,[2/3,-1/3,-1/3] = (\lambda_3 +
\lambda_8/\sqrt{3}\,)/2$\\
\makebox[6.4mm]{} and the $\bar{n}n = (\bar{u}u + \bar{d}d)/\sqrt{2}$ analogue
$Q_{\bar{n}n} = $Diag$\,[1/\sqrt{2},1/\sqrt{2},0] = (\lambda_0 +
\lambda_8/\sqrt{2}\,)/\sqrt{3}\,$.}.
If, instead,  the $f_0 (980)$ is a pure $\bar{s}s$ state, the
$f_0\rightarrow 2\,\gamma$ amplitude magnitude becomes \cite{DLS99}
$\alpha \, N_c \,\, g_{{}_{f_0\, SS}}\, / (9\,\pi\,m_s) \simeq 0.81\cdot
10^{-2}$ GeV${}^{-1}$, using $g_{{f_0\,SS}} = \sqrt{2}\;2\pi /\sqrt{3}$ and
constituent strange quark mass \cite{DS98,S82} $m_s = 490$ MeV $\simeq 1.44 \;
\hat{m}$ (from Ref.~\cite{DS98}, $F_K/F_\pi = (\hat{m} + m_s) / (2\, \hat{m} )
\simeq 1.22$) with the constituent nonstrange mass \mbox{$\hat{m}\simeq
340$ MeV}. This value lies reasonably close the
observed amplitude in Eq.~(\ref{eq1}).\footnote{Without changes, we could of
course also use the identity $\sqrt{2}\,\alpha\,N_c\,$Tr$\,[Q^{2\,} \,
Q_{\bar{s}s}] \, /\,(\pi\,F_{\bar{s}s}) = \sqrt{2}\;\alpha\,N_c\,/\,(9 \,
\pi\,F_{\bar{s}s})\simeq 0.81\cdot 10^{-2}$ GeV${}^{-1}$, with
$F_{\bar{s}s} =\sqrt{3}\, m_s /(2 \, \pi) = 135.1\;$MeV $\simeq 1.2 \, F_K \,
\simeq 2F_K-F_{\pi}\simeq \sqrt{2} \, F_\pi$ and $Q_{\bar{s}s} = $Diag$\,[0,0,1] = (\lambda_0/\sqrt{2}
- \lambda_8 \,)/\sqrt{3}\,$. The use of \cite{PDG2000} $F_K=f_{K^+}/\sqrt{2}
=(113.00\pm 1.04)$ MeV instead of $F_{\bar{s}s}$ would bring us even closer to
the data, as $\sqrt{2}\;\alpha\,N_c\,/\,(9 \, \pi\,F_{K})\simeq
0.972\cdot 10^{-2}$ GeV${}^{-1}$.}  
However, at this point we should note that the quark-loop result for the
two-photon decay rate is very sensitive to a possible $\bar{n}n$ admixture
in the \fn, due to an enhancement factor of 25 of the $\bar{n}n$ component
with respect to the $\bar{s}s$ component. This factor comes from the electric
charge of the quarks, yielding $((\frac{2}{3})^2+(\frac{1}{3})^2)^2$  for the
nonstrange isoscalar $\frac{1}{\sqrt{2}}(\bar{u}u+\bar{d}d)$, and
$(\frac{1}{3})^4$ for the strange isoscalar.

Therefore, rather than involving the model-dependent quark coupling and
constituent quark masses as above, we instead consider a combination of the
decay chains $f_0 \rightarrow K^+K^-\rightarrow 2\,\gamma$ and $f_0 \rightarrow
\pi^+\pi^-\rightarrow 2\,\gamma$ \cite{DEMSB94,DLS99,BBIS00,LN99}. According to
Refs.\ \cite{DEMSB94,LN99}, the kaon loop is suppressed by 10\% due to a,
so far experimentally unconfirmed, scalar $\kappa (900)$. (However, very recent
results from the E791 collaboration present preliminary evidence for a light
$\kappa$ (see e-print in Ref.~\cite{G00}), which would confirm the prediction
\cite{S82,BRMDRR86} of such a state.) In order to proceed, we have to
remind the reader to the standard mixing scheme between the ``physical'' states
($|\sigma (600)>$ and $|f_0(980)>$), and the nonstrange and strange basis
states $|\bar{n}n>$ and $|\bar{s}s>$, i.e.,
\begin{eqnarray} |\sigma(600)> & = & \cos \phi_s \; |\bar{n}n> \, - \,
\sin \phi_s \; |\bar{s}s> \, , \nonumber \\
|f_0(980)> & = & \sin \phi_s \; |\bar{n}n> \; + \, \cos \phi_s \;
|\bar{s}s> \, . \label{eq2}
\end{eqnarray}
With quadratic mass mixing, one can define for the states $|\bar{n}n>$ and
$|\bar{s}s>$ the nonstrange and strange mass parameters $m_{\bar{n}n}$ and
$m_{\bar{s}s}$ by \cite{DLS99} as
\begin{eqnarray} m^2_{\bar{n}n} & = & \cos^2 \phi_s \;  m^2_{\sigma} \, +
\, \sin^2 \phi_s \;  m^2_{f_0} \; = \; \Big( (646 \pm 10)\; \makebox{MeV}
\Big)^2 \, , \nonumber \\
m^2_{\bar{s}s} & = & \sin^2 \phi_s \;  m^2_{\sigma} \, + \, \cos^2 \phi_s \;
m^2_{f_0} \; = \; \Big( (950 \pm 11)\; \makebox{MeV} \Big)^2 \, . \label{eq3} 
\end{eqnarray}
Throughout this paper we choose a mixing angle of \footnote{The sign of the
mixing angle, which cannot be identified from a quadratic mass mixing scheme,
has still to be determined from theoretical consistency arguments, as it 
has a strong influence on the interference terms in the present work.}
$\phi_s \simeq 18^\circ \pm 2^\circ$ \cite{S82,S84a,DS98,DLS99} or
$\phi_s \simeq -\, ( 18^\circ \pm 2^\circ)$ \cite{LN99}, and assume the
scalar-meson masses
to be $m_{f_0(980)} = (980 \pm 10)$~MeV~\cite{PDG2000} and $m_{\sigma(600)}=
600$~MeV.
Since the interaction Lagrangians between the $f_0$ and the pseudoscalars
$\pi^\pm$ and $K^\pm$ are proportional to $f_0$, the Lagrangians can, within
the same mixing scheme, be simultaneously reexpressed in terms of nonstrange
and strange fields, i.e.,
\begin{eqnarray} \lefteqn{{\cal L} \, (f_0 \pi\pi) + {\cal L} \, (f_0 K K) =}
\nonumber \\
 & = & \sin \phi_s \; \Big({\cal L} \, (\bar{n}n \, \pi\pi) + {\cal L} \,
(\bar{n}n \, KK) \Big)  \; + \, \cos \phi_s \; \Big({\cal L} \, (\bar{s}s \,
\pi\pi) + {\cal L} \, (\bar{s}s \, KK) \Big) \label{eq4} 
\end{eqnarray}
Within the usual nonet, that is, the U(3) picture, the scalar (S) and
pseudoscalar
(P) fields are proportional to linear combinations of the Gell-Mann matrices
$\lambda_0,\lambda_1, \ldots, \lambda_8$ ($\lambda_0$ denotes here $\sqrt{2/3}
\;\, 1_3$ with $1_3$ being the 3-dimensional unit matrix), denoted by $Q_S$ and
$Q_P$, respectively. From the quark content of the corresponding mesonic
systems, it is easy to derive

\begin{eqnarray} \bar{n}\,n = \frac{1}{\sqrt{2}} \; (\bar{u}u + \bar{d}d) &
\Rightarrow  & Q_{ \bar{n}\,n} = \frac{1}{\sqrt{3}} \, \left(\lambda_0 +
\frac{1}{\sqrt{2}}\;\lambda_8\right) \, , \nonumber \\ 
\bar{s}\,s & \Rightarrow  & Q_{ \bar{s}\,s} = \frac{1}{\sqrt{3}} \, \left(
\frac{1}{\sqrt{2}}\; \lambda_0 - \, \lambda_8\right) \, , \nonumber \\ 
 & & \nonumber \\
\pi^+ = \bar{d} u\;,\; \pi^- = \bar{u} d & \Rightarrow & Q_{\pi^\pm} =
\frac{1}{2} \, \left(\lambda_1 \pm i\;\lambda_2\right) \, , \nonumber \\ 
K^+ = \bar{s} u\;,\; K^- = \bar{u} s & \Rightarrow & Q_{K^\pm} = \frac{1}{2}
\, \left(\lambda_4 \pm i\;\lambda_5\right) \, . 
\end{eqnarray}

In the linear $\sigma$ model (LSM), the interaction Lagrangian ${\cal L} \,
(S \, P_1 \, P_2 )$ is proportional to the flavor trace
Tr$\;(Q_S\,\{Q_{P_1},Q_{P_2}\})$, and so are the corresponding coupling
constants. It should be mentioned that the charge of a mesonic system $\phi$ is
determined by Tr$\;(Q\,[Q_{\phi},Q^T_{\phi}])$. Thus, we derive for the
relevant channels under consideration, i.e., $\bar{n}n \rightarrow \pi\pi$,
$\bar{n}n \rightarrow KK$, $\bar{s}s \rightarrow \pi\pi$, and $\bar{s}s
\rightarrow KK$:
\begin{eqnarray} d_{\,\bar{n}n\,\pi^+\pi^-} \; = & \displaystyle
\frac{1}{\sqrt{2}} \; \mbox{Tr}\;(Q_{\bar{n}n}\,\{Q_{\pi^+},Q_{\pi^-}\}) & =
\; 1 \, , \nonumber \\
 d_{\,\bar{n}n \,K^+K^-} \; = & \displaystyle \frac{1}{\sqrt{2}} \;
\mbox{Tr}\;(Q_{\bar{n}n}\,\{Q_{K^+},Q_{K^-}\}) & = \; \frac{1}{2} \, , 
\nonumber \\
d_{\,\bar{s}s \,\pi^+\pi^-} \; = & \displaystyle \frac{1}{\sqrt{2}} \;
\mbox{Tr}\;(Q_{\bar{s}s}\,\{Q_{\pi^+},Q_{\pi^-}\}) & = \; 0 \, , \nonumber \\
d_{\,\bar{s}s \, K^+K^-} \; = & \displaystyle  \frac{1}{\sqrt{2}} \;
\mbox{Tr}\;(Q_{\bar{s}s}\,\{Q_{K^+},Q_{K^-}\}) & = \; \frac{1}{\sqrt{2}} \, .
\label{eq5} \end{eqnarray}  
The corresponding equivalent symmetric structure constants
$d_{\,\bar{n}n\,33}$, $d_{\,\bar{n}n\,K^0K^0}$, $d_{\,\bar{s}s\,33}$,
$d_{\,\bar{s}s\,K^0K^0}$, with $d_{abc} = \mbox{Tr} (\lambda_a \{
\lambda_b,\lambda_c\}) /4$, for two neutral pseudoscalars in the final state
have already been derived in Ref.~\cite{DS98}.
In accordance with the $\sigma$-model results, we determine the corresponding
SU(3) couplings for $\phi_s \simeq + \, (18^\circ\pm 2^\circ)$ and $\phi_s
\simeq - \, (18^\circ\pm 2^\circ)$ as 
\begin{eqnarray} g_{\bar{n}n\,\pi\pi}^{\,\prime} & = &
d_{\,\bar{n}n\,\pi^+\pi^-} \;\, \frac{m^2_{\bar{n}n} -
m^2_{\pi^\pm}}{2\,F_\pi} 
 = \frac{\cos^2 \phi_s \;  m^2_{\sigma} \, + \, \sin^2 \phi_s \;  m^2_{f_0}
- m^2_{\pi^\pm}}{2\,F_\pi} \nonumber \\
 & & \nonumber \\
 & = &  (2.152 \pm 0.068) \; \mbox{GeV} \, , \nonumber \\ 
 & & \nonumber \\
g_{\bar{n}n\,KK}^{\,\prime} & = & d_{\,\bar{n}n \,K^+K^-} \;\,
\frac{m^2_{\bar{n}n} - m^2_{K^\pm}}{F_K}
 = \frac{\cos^2 \phi_s \;  m^2_{\sigma} \, + \, \sin^2 \phi_s \;
 m^2_{f_0} - m^2_{K^\pm}}{2\, F_K} \nonumber \\
 & & \nonumber \\
 & = & (0.768 \pm 0.056) \; \mbox{GeV} \, , \nonumber \\ 
 & & \nonumber \\
g_{\bar{s}s\,\pi\pi}^{\,\prime} & = & d_{\,\bar{s}s \,\pi^+\pi^-} \;\,
\frac{m^2_{\bar{s}s} - m^2_{\pi^\pm}}{2\,F_\pi} \; = \; 0 \, , \nonumber \\ 
 & & \nonumber \\
g_{\bar{s}s\,KK}^{\,\prime} & = & d_{\,\bar{s}s \, K^+K^-} \;\,
\frac{m^2_{\bar{s}s} - m^2_{K^\pm}}{F_K} 
 = \frac{\sin^2 \phi_s \;  m^2_{\sigma} \, + \, \cos^2 \phi_s \;  m^2_{f_0}
- m^2_{K^\pm}}{\sqrt{2} \; F_K} \nonumber \\
 & & \nonumber \\
 & = & (4.126 \pm 0.141) \; \mbox{GeV} \, , \label{eq6}
\end{eqnarray}
yielding for $\phi_s \simeq + \, (18^\circ\pm 2^\circ)$
\begin{eqnarray} (\sin\, \phi_s\;\; g_{\bar{n}n\,\pi\pi}^
{\,\prime}\,\;+ \;\cos\, \phi_s\; g_{\bar{s}s\,\pi\pi}^{\,\prime}) & = &
(0.665 \pm 0.093) \; \mbox{GeV} \, , \nonumber \\
(\sin\, \phi_s\; g_{\bar{n}n\,KK}^{\,\prime}\,+\cos\, \phi_s\;
g_{\bar{s}s\,KK}^{\,\prime}) & = & (4.162 \pm 0.138) \; \mbox{GeV} \, ,
\end{eqnarray}
and for $\phi_s \simeq - \, (18^\circ\pm 2^\circ)$
\begin{eqnarray} (\sin\, \phi_s\;\; g_{\bar{n}n\,\pi\pi}^
{\,\prime}\,\;+\;\cos\, \phi_s\; g_{\bar{s}s\,\pi\pi}^{\,\prime}) & = &
 (-\,0.665\pm 0.093) \; \mbox{GeV} \, , \nonumber \\
(\sin\, \phi_s\; g_{\bar{n}n\,KK}^{\,\prime}\,+\cos\, \phi_s\;
g_{\bar{s}s\,KK}^{\,\prime}) & = & (3.687 \pm 0.194) \; \mbox{GeV} \, .
\end{eqnarray}

In order to compute these numbers, we used $F_\pi \simeq (92.42 \pm 0.27)\;
\mbox{MeV}$, $F_K \simeq (113.00 \pm 1.04)\; \mbox{MeV}$, i.e. $F_K / F_\pi
\simeq 1.22$.
Putting all this together, we obtain for the pion- and kaon-loop amplitudes
 \cite{DEMSB94}  
\begin{eqnarray} \lefteqn{M_{\,\mbox{\small $\pi$-loop}} =
\frac{2\,\alpha\,(\sin\, \phi_s\; g_{\bar{n}n\,\pi\pi}^{\,\prime}\,+\cos\,
\phi_s\; g_{\bar{s}s\,\pi\pi}^{\,\prime})}{\pi\,m^2_{f_0}} \,\left[ -\,
\frac{1}{2} + \xi_\pi \, I(\xi_\pi )  \right] } \nonumber \\
  & = & (-0.177 \pm 0.025 + i\, (+\,0.079 \pm 0.012)) \cdot 10^{-2}\;
\mbox{GeV}{}^{-1} \quad \mbox{for} \;\; \phi_s \simeq + \,
(18^\circ\pm 2^\circ) \nonumber  \\
  & = & (+0.177 \pm 0.025 + i\, (-\,0.079 \pm 0.012)) \cdot 10^{-2}\;
\mbox{GeV}{}^{-1} \quad \mbox{for} \;\; \phi_s \simeq - \,
(18^\circ\pm 2^\circ) \, , \nonumber  \\
 & & \nonumber \\
\lefteqn{M_{\,\mbox{\small K-loop}} = \frac{2\,\alpha\,(\sin\, \phi_s\;
g_{\bar{n}n\,KK}^{\,\prime}\,+\cos\, \phi_s\; g_{\bar{s}s\,KK}^
{\,\prime})}{\pi\,m^2_{f_0}} \,\left[ -\, \frac{1}{2} + \xi_K \, I(\xi_K ) 
\right]} \nonumber \\
  & = & (1.138 \pm 0.254) \cdot 10^{-2}\; \mbox{GeV}{}^{-1} \quad \mbox{for}
\;\; \phi_s \simeq + \, (18^\circ\pm 2^\circ) \nonumber \\
  & = & (1.008 \pm 0.229) \cdot 10^{-2}\; \mbox{GeV}{}^{-1} \quad \mbox{for}
\;\; \phi_s \simeq - \, (18^\circ\pm 2^\circ) \, , \nonumber \\
 & & \nonumber \\
\lefteqn{M_{\,\mbox{\small $\pi$-loop}} + M_{\,\mbox{\small K-loop}} = }
\nonumber \\ 
& = & (0.960 \pm 0.255 + i\, (+\,0.079 \pm 0.012)) \cdot 10^{-2}\;
\mbox{GeV}{}^{-1} \quad \mbox{for} \;\; \phi_s \simeq + \,
(18^\circ\pm 2^\circ) \nonumber \\
 & = &
(1.185 \pm 0.230 + i\, (-\,0.079 \pm 0.012)) \cdot 10^{-2}\;
\mbox{GeV}{}^{-1} \quad \mbox{for} \;\; \phi_s \simeq - \, (18^\circ\pm
 2^\circ) \, , \nonumber \\
 & & \nonumber \\
\lefteqn{|M_{\,\mbox{\small $\pi$-loop}} + M_{\,\mbox{\small K-loop}}| = }
\nonumber \\ 
& = & (0.964 \pm 0.255) \cdot 10^{-2}\; \mbox{GeV}{}^{-1} \quad \mbox{for} \;\;
\phi_s \simeq + \, (18^\circ\pm 2^\circ) \nonumber \\
 & = &
(1.188 \pm 0.230) \cdot 10^{-2}\; \mbox{GeV}{}^{-1} \quad \mbox{for} \;\;
\phi_s \simeq - \, (18^\circ\pm 2^\circ) \, .
 \label{eq7}
\end{eqnarray} 
As $\xi_\pi = m^2_{\pi^+} / m^2_{f_0(980)} = 0.02028 \pm 0.00042 < 1/4$,
the value of the pion-loop integral is obtained from (see also p.\ 230, 422 in
Ref.~\cite{D92})
\begin{eqnarray} I(\xi_\pi) & = & \int_0^1dy \int_0^1dx\; \frac{y}{\xi_\pi -xy
 \, (1-y)} \; = \; 2 \left[ \frac{\pi}{2} + i \, \ln \left(
 \sqrt{\frac{1}{4\,\xi_\pi} } + \sqrt{\frac{1}{4\,\xi_\pi} -1 }\; \right)
 \right]^2 \nonumber \\
 & & \nonumber \\
 & = &  \frac{\pi^2}{2} - 2 \, \ln^{\,2} \left[ \sqrt{\frac{1}{4\,\xi_\pi} } +
 \sqrt{\frac{1}{4\,\xi_\pi} -1 }\; \right] + 2 \, \pi \, i \, \ln \left[
 \sqrt{\frac{1}{4\,\xi_\pi} } + \sqrt{\frac{1}{4\,\xi_\pi} -1 }\; \right]
 \nonumber \\
 & & \nonumber \\
 & = & -\, 2.500 \pm 0.083 + i\, (12.114 \pm 0.067) \, , \nonumber 
\label{eq8}
\end{eqnarray}
while, as $\xi_K = m^2_{K^+} / m^2_{f_0(980)} = 0.2538 \pm 0.0052 > 1/4$,
the kaon loop follows from
\begin{equation} I(\xi_K) \; = \; \int_0^1dy \int_0^1dx\; \frac{y}{\xi_K -xy \,
 (1-y)} \; = \; 2 \left[ \arcsin \sqrt{\frac{1}{4\,\xi_K} } \; \right]^2 =
 4.197 \pm 0.482  \, ,\label{eq9}
\end{equation}
yielding, respectively,
\begin{eqnarray} 
 -\, \frac{1}{2} + \xi_\pi \, I(\xi_\pi ) \; & = & -\,0.5507 \pm 0.0020 +
 i\, (0.2457 \pm 0.0037) \, , \nonumber \\
 -\, \frac{1}{2} + \xi_K \, I(\xi_K ) & = & 0.5651 \pm 0.1242 \, .
\end{eqnarray}
Reducing the kaon-loop amplitude in Eq.\ (\ref{eq7}) by 10 \% (owing to the
scalar $\kappa (900)$ loop), but leaving the value of its error unaltered,
predicts $(0.85 \pm 0.26) \cdot 10^{-2}\; \mbox{GeV}{}^{-1}$ ($\phi_s \simeq +
 \, (18^\circ\pm 2^\circ)$) or $(1.09 \pm 0.23) \cdot 10^{-2}\;
 \mbox{GeV}{}^{-1}$ ($\phi_s \simeq  \, -\,(18^\circ\pm 2^\circ)$) for the
modulus of the $f_0(980)\rightarrow 2\,\gamma$ amplitude, reasonably near the
 data~\cite{PDG2000} in Eq.~(\ref{eq1}). Therefore, whether we employ quark
loops or instead $\pi$ and $K$ loops as in Eq.~(\ref{eq7}), it is clear that
the $f_0(980)\rightarrow 2\,\gamma$ amplitude can only be understood, if the
$f_0(980)$ is mostly $\bar{s}s$.\footnote{Surely, the error bars of the
presented analysis rely strongly on the assumption that we choose a sharp
$\sigma$-meson mass $m_\sigma = 600$ MeV, without any uncertainty.} This is
the same conclusion as obtained, more easily, from the weak decay
$D^+_s\rightarrow \pi^+ f_0(980)$ in Eq.~(1).

Similar conclusions for the flavor content of the \fn\ can be found
in Refs.~\cite{Fulvia,DGNPT00}. Furthermore, in Ref.~\cite{AABN99} two
possibilities are indicated, either dominantly $\bar{s}s$, or flavor octet,
which is dominantly $\bar{s}s$ as well.
 
\subsection{The decay $f_0(1370) \rightarrow 2\,\gamma$}
Now we study the process $f_0(1370)\rightarrow 2\,\gamma$, using the same
techniques as above. In the meson listings of the Particle Data Group
\cite{PDG2000}, two values are given for the two-photon partial width of the
\ft, i.e., $(3.8 \pm 1.5)$ keV and $(5.4 \pm 2.3)$ keV, from Refs.~\cite{BP98}
and \cite{MP90}, respectively. In these analyses, the $2\gamma$ coupling is
determined from the $S$-wave $\gamma\gamma\to\pi\pi$ cross section in the
energy region under the $f_2(1270)$. However, the peaking of this cross section
above 1 GeV is explained by the authors as a consequence of a low-mass-scalar
suppression due to gauge invariance (see also Ref.~\cite{P00}), pushing the
corresponding distribution towards the high-mass end of the \fsigma, rather
than as a signal of the \ft. For the purpose of our present study, we abide by
the current PDG interpretation favoring the \ft, but keeping in mind that the
experimental situation is anything but settled. Furthermore, we average the two
data on the two-photon partial width, providing us a, albeit preliminary,
theoretical value of $(4.6 \pm 2.8)$ keV,
with amplitude given by (using $m_{f_0(1370)}=(1370\pm 170)$ MeV)
\begin{equation}\Gamma (f_0\rightarrow 2\,\gamma) = \frac{m_{\,f^{\,0}}^{\,3}
\, |M|^2}{64\,\pi}\, , \;\; \mbox{or} \quad |M(f_0(1370)\rightarrow 2\,\gamma)|
 = (1.90 \pm 0.68)\cdot 10^{-2}\;\mbox{GeV}^{-1} \, .\label{eq10}
\end{equation}

In order to apply again a meson-loop approach, we develop once more a
meson-mixing scheme, namely
\begin{eqnarray} |f_0(1370)> & = & \cos \phi^{\,\,\prime}_s \; |\bar{n}n> \, -
 \, \sin \phi^{\,\,\prime}_s \; |\bar{s}s> \nonumber \\
|f_0(1500)> & = & \sin \phi^{\,\,\prime}_s \; |\bar{n}n> \; + \,
 \cos \phi^{\,\,\prime}_s \; |\bar{s}s> \, .
\end{eqnarray}
Again we define, using quadratic mass mixing with respect to the states
$|\bar{n}n>$ and $|\bar{s}s>$, the nonstrange and strange mass parameters
$m^{\prime}_{\bar{n}n}$ and $m^{\prime}_{\bar{s}s}$ by
\begin{eqnarray} m^{\prime\,2}_{\bar{n}n} & = & \cos^2 \phi^{\,\,\prime}_s \;
 m^2_{f_0(1370)} \, + \, \sin^2 \phi^{\,\,\prime}_s \;  m^2_{f_0(1500)} \, ,
 \nonumber \\
m^{\prime\,2}_{\bar{s}s} & = & \sin^2 \phi^{\,\,\prime}_s \;  m^2_{f_0(1370)}
 \, + \, \cos^2 \phi^{\,\,\prime}_s \;  m^2_{f_0(1500)} \, .
\end{eqnarray}
Consequently, we use the couplings
\begin{eqnarray} g_{\bar{n}n\,\pi\pi}^{\,\prime} & = &
 d_{\,\bar{n}n\,\pi^+\pi^-} \;\, \frac{m^{\prime\,2}_{\bar{n}n} -
 m^2_{\pi^\pm}}{2\,F_\pi} 
 = \frac{\cos^2 \phi^{\,\,\prime}_s \;  m^2_{f_0(1370)} \, +
 \, \sin^2 \phi^{\,\,\prime}_s \;  m^2_{f_0(1500)} - m^2_{\pi^\pm}}{2\,F_\pi}
 \nonumber \\
 & & \nonumber \\
 & = & (10.05 \pm 2.53) \;
\mbox{GeV}{} \quad \mbox{for} \;\; \phi^{\,\,\prime}_s = 0^\circ \, , \nonumber
 \\ & = & (10.24 \pm 2.29) \;
\mbox{GeV}{} \quad \mbox{for} \;\; \phi^{\,\,\prime}_s \simeq \pm \,
 (18^\circ\pm 2^\circ) \, , \nonumber \\
 & & \nonumber \\
g_{\bar{n}n\,KK}^{\,\prime} & = & d_{\,\bar{n}n \,K^+K^-} \;\,
 \frac{m^{\prime\,2}_{\bar{n}n} - m^2_{K^\pm}}{F_K}
 = \frac{\cos^2 \phi^{\,\,\prime}_s \;  m^2_{f_0(1370)} \, +
 \, \sin^2 \phi^{\,\,\prime}_s \;  m^2_{f_0(1500)} - m^2_{K^\pm}}{2\, F_K}
 \nonumber \\
 & & \nonumber \\
 & = & (7.23 \pm 2.07) \;
\mbox{GeV}{} \quad \mbox{for} \;\; \phi^{\,\,\prime}_s = 0^\circ \, , \nonumber
 \\ & = & (7.38 \pm 1.87) \;
\mbox{GeV}{} \quad \mbox{for} \;\; \phi^{\,\,\prime}_s \simeq \pm \,
 (18^\circ\pm 2^\circ) \, , \nonumber \\
 & & \nonumber \\
g_{\bar{s}s\,\pi\pi}^{\,\prime} & = & d_{\,\bar{s}s \,\pi^+\pi^-} \;\,
 \frac{m^{\prime\,2}_{\bar{s}s} - m^2_{\pi^\pm}}{2\,F_\pi} \; = \; 0
 \nonumber \\ 
 & & \nonumber \\
g_{\bar{s}s\,KK}^{\,\prime} & = & d_{\,\bar{s}s \, K^+K^-} \;\,
 \frac{m^{\prime\,2}_{\bar{s}s} - m^2_{K^\pm}}{F_K} 
 = \frac{\sin^2 \phi^{\,\,\prime}_s \;  m^2_{f_0(1370)} \, +
 \, \cos^2 \phi^{\,\,\prime}_s \;  m^2_{f_0(1500)} - m^2_{K^\pm}}{\sqrt{2}
 \; F_K} \nonumber \\
 & & \nonumber \\
 & = & (12.56 \pm 0.23) \;
\mbox{GeV}{} \quad \mbox{for} \;\; \phi^{\,\,\prime}_s = 0^\circ \, , \nonumber
 \\ & = & (12.33 \pm 0.35) \;
\mbox{GeV}{} \quad \mbox{for} \;\; \phi^{\,\,\prime}_s \simeq \pm \,
 (18^\circ\pm 2^\circ) \, ,
\end{eqnarray}
yielding, respectively, 
\begin{eqnarray} (\cos\,
 \phi^{\,\,\prime}_s\; g_{\bar{n}n\,\pi\pi}^{\,\prime}\; \,- \sin\,
 \phi^{\,\,\prime}_s\; g_{\bar{s}s\,\pi\pi}^{\,\prime}) \;\; & = &
 (10.05 \pm 2.53) \;
\mbox{GeV}{} \quad \mbox{for} \;\; \phi^{\,\,\prime}_s = 0^\circ \, , \nonumber
 \\ & = & (9.74 \pm 2.17) \;
\mbox{GeV}{} \quad \;\:\mbox{for} \;\; \phi^{\,\,\prime}_s \simeq + \,
 (18^\circ\pm 2^\circ) \, , 
\nonumber \\
 & = & (9.74 \pm 2.17) \;
\mbox{GeV}{} \quad \;\:\mbox{for} \;\; \phi^{\,\,\prime}_s \simeq - \,
 (18^\circ\pm 2^\circ) \, , 
\nonumber \\
(\cos\,
 \phi^{\,\,\prime}_s\; g_{\bar{n}n\,KK}^{\,\prime}\,-\sin\,
 \phi^{\,\,\prime}_s\; g_{\bar{s}s\,KK}^{\,\prime}) & = &
 (7.23 \pm 2.07)
\; \mbox{GeV}{} \quad \;\:\mbox{for} \;\; \phi^{\,\,\prime}_s = 0^\circ \, ,
 \nonumber \\ & = & (3.21 \pm 1.75) \;
\mbox{GeV}{} \quad \;\:\mbox{for} \;\; \phi^{\,\,\prime}_s \simeq + \,
 (18^\circ\pm 2^\circ) \, , 
 \nonumber \\ & = & (10.83 \pm 1.90) \;
\mbox{GeV}{} \quad \mbox{for} \;\; \phi^{\,\,\prime}_s \simeq - \,
 (18^\circ\pm 2^\circ) \, ,
\end{eqnarray}
to determine the pion- and kaon-loop amplitudes
\begin{eqnarray} M_{\,\mbox{\small $\pi$-loop}} & = & \frac{2\,\alpha\,(\cos\,
 \phi^{\,\,\prime}_s\; g_{\bar{n}n\,\pi\pi}^{\,\prime}\,-\sin\,
 \phi^{\,\,\prime}_s\; g_{\bar{s}s\,\pi\pi}^{\,\prime})}{\pi\,m^2_{f_0(1370)}}
 \,\left[ -\, \frac{1}{2} + \xi_\pi \, I(\xi_\pi )  \right]  \, , \nonumber \\
 & & \nonumber \\
 M_{\,\mbox{\small K-loop}} & = & \frac{2\,\alpha\,(\cos\,
 \phi^{\,\,\prime}_s\; g_{\bar{n}n\,KK}^{\,\prime}\,-\sin\,
 \phi^{\,\,\prime}_s\; g_{\bar{s}s\,KK}^{\,\prime})}{\pi\,m^2_{f_0(1370)}}
 \,\left[ -\, \frac{1}{2} + \xi_K \, I(\xi_K )  \right] \, .
\end{eqnarray}
 Using $\xi_\pi = m^2_{\pi^+} / m^2_{f_0(1370)} = 0.0104 \pm 0.0026 < 1/4$
 and $\xi_K = m^2_{K^+} / m^2_{f_0(1370)} = 0.1299 \pm 0.0323 < 1/4$, we
obtain \cite{DEMSB94} (see also p.\ 230, 422 in Ref.~\cite{D92})
\begin{eqnarray} 
I(\xi_\pi) & = &  \frac{\pi^2}{2} - 2 \, \ln^{\,2} \left[
 \sqrt{\frac{1}{4\,\xi_\pi} } + \sqrt{\frac{1}{4\,\xi_\pi} -1 }\; \right] +
 2 \, \pi \, i \, \ln \left[ \sqrt{\frac{1}{4\,\xi_\pi} } +
 \sqrt{\frac{1}{4\,\xi_\pi} -1 }\; \right] \nonumber \\
& & \nonumber \\
 & = & -5.40 \pm 1.16 + i\,(14.28 \pm 0.80) \, , \nonumber \\ 
& & \nonumber \\
I(\xi_K) & = & \frac{\pi^2}{2} - 2 \, \ln^{\,2} \left[
 \sqrt{\frac{1}{4\,\xi_K} } + \sqrt{\frac{1}{4\,\xi_K} -1 }\; \right] + 2 \,
 \pi \, i \, \ln \left[ \sqrt{\frac{1}{4\,\xi_K} } +
 \sqrt{\frac{1}{4\,\xi_K} -1 }\; \right] \nonumber \\
 & & \nonumber \\
 & = & 3.48 \pm 0.62  + i\,(5.37 \pm 1.13) \, ,
\label{eq13}
\end{eqnarray}
yielding, respectively, 
\begin{eqnarray} 
 -\, \frac{1}{2} + \xi_\pi \, I(\xi_\pi ) \; & = & -\, 0.556 \pm 0.002
 + i\,(0.148 \pm 0.029) \, , \nonumber \\
 -\, \frac{1}{2} + \xi_K \, I(\xi_K ) & = & -\, 0.049 \pm 0.192
  + i\,(0.697 \pm 0.027) \, .
\end{eqnarray}
Combining all the previous results, we arrive at
\begin{eqnarray} \lefteqn{M_{\,\mbox{\small $\pi$-loop}} =
 } \nonumber \\
 & = & ( -\,1.383 \pm 0.008 + i\, (0.369 \pm 0.071)) \cdot
 10^{-2}\;
\mbox{GeV}{}^{-1} \quad\;\; \mbox{for} \;\; \phi^{\,\,\prime}_s = 0^\circ \, ,
\nonumber  \\
  & = & ( - \, 1.341 \pm 0.037 + i\, (0.357 \pm 0.070)) \cdot
 10^{-2}\;
\mbox{GeV}{}^{-1} \quad\;\; \mbox{for} \;\; \phi^{\,\,\prime}_s \simeq + \,
(18^\circ\pm 2^\circ) \, , \nonumber  \\
  & = & ( - \, 1.341 \pm 0.037 + i\, (0.357 \pm 0.070)) \cdot
 10^{-2}\;
\mbox{GeV}{}^{-1} \quad\;\; \mbox{for} \;\; \phi^{\,\,\prime}_s \simeq - \,
(18^\circ\pm 2^\circ) \, , \nonumber  \\
\lefteqn{M_{\,\mbox{\small K-loop}} = } \nonumber \\
  & = & (-\,0.087 \pm 0.343 + i\, (1.247 \pm 0.068)) \cdot
 10^{-2}\;
\mbox{GeV}{}^{-1} \quad\;\; \mbox{for} \;\; \phi^{\,\,\prime}_s = 0^\circ \, , 
\nonumber  \\
  & = & (-\,0.039 \pm 0.153 + i\, (0.554 \pm 0.173) ) \cdot
 10^{-2}\; \mbox{GeV}{}^{-1} \quad\;\; \mbox{for}
\;\; \phi^{\,\,\prime}_s \simeq + \, (18^\circ\pm 2^\circ) \, , \nonumber \\
  & = & (-\,0.131 \pm 0.514 + i\, (1.869 \pm 0.173) ) \cdot
 10^{-2}\; \mbox{GeV}{}^{-1} \quad\;\; \mbox{for}
\;\; \phi^{\,\,\prime}_s \simeq - \, (18^\circ\pm 2^\circ) \, , \nonumber \\
\lefteqn{M_{\,\mbox{\small $\pi$-loop}} + M_{\,\mbox{\small K-loop}} = }
\nonumber \\ 
  & = & (-\,1.470 \pm 0.343 + i\, (1.615 \pm 0.099)) \cdot
 10^{-2}\;
\mbox{GeV}{}^{-1} \quad\;\; \mbox{for} \;\; \phi^{\,\,\prime}_s = 0^\circ \, ,
\nonumber  \\
 & = & (-\,1.379 \pm 0.157 + i\, (0.912 \pm 0.187)) \cdot
 10^{-2}\;
\mbox{GeV}{}^{-1} \quad\;\; \mbox{for} \;\; \phi^{\,\,\prime}_s \simeq + \,
(18^\circ\pm 2^\circ) \, , \nonumber \\
 & = &
 (-\,1.471 \pm 0.515 + i\, (2.226 \pm 0.186)) \cdot
 10^{-2}\;
\mbox{GeV}{}^{-1} \quad \;\;\mbox{for} \;\; \phi^{\,\,\prime}_s
 \simeq - \, (18^\circ\pm 2^\circ) \, , \nonumber \\
\lefteqn{|M_{\,\mbox{\small $\pi$-loop}} + M_{\,\mbox{\small K-loop}}| = }
\nonumber \\ 
  & = & (2.184 \pm 0.242) \cdot 10^{-2}\;
\mbox{GeV}{}^{-1} \quad\;\; \mbox{for} \;\; \phi^{\,\,\prime}_s = 0^\circ \, ,
\nonumber  \\
 & = & (1.653 \pm 0.167) \cdot 10^{-2}\; \mbox{GeV}{}^{-1} \quad\;\;
 \mbox{for} \;\;
\phi^{\,\,\prime}_s \simeq + \, (18^\circ\pm 2^\circ) \, , \nonumber \\
 & = &
 (2.668 \pm 0.324) \cdot 10^{-2}\; \mbox{GeV}{}^{-1} \quad\;\;
 \mbox{for} \;\;
\phi^{\,\,\prime}_s \simeq - \, (18^\circ\pm 2^\circ) \, .
\end{eqnarray} 

If we again reduce the kaon-loop amplitude by 10\% owing to the $\kappa$(900),
and assume for the moment that the \ft\ is purely $\bar{n}n$, we get for the
modulus of the decay amplitude the value $(2.09 \pm 0.25) \cdot 10^{-2}$
GeV$^{-1}$, in good agreement with the experimental result in Eq.~(\ref{eq10}).
Taking instead a mixing angle of $\phi^{\,\,\prime}_s = 18^\circ\pm 2^\circ$
produces an amplitude 
value of $(1.62 \pm 0.17) \cdot 10^{-2}$ GeV$^{-1}$, also well within the 
experimental error bars. On the other hand, choosing a negative mixing angle of
$\phi^{\,\,\prime}_s = -18^\circ\pm 2^\circ$ gives rise to a somewhat too large
amplitude, albeit still compatible with the experimentally allowed range of
values, namely $(2.51 \pm 0.34) \cdot 10^{-2}$ GeV$^{-1}$. So a
positive mixing angle seems to be clearly favored. Further increasing a
positive $\phi^{\,\,\prime}_s$ from $+18^\circ$ will yield smaller and smaller
amplitudes, until at about $60^\circ$ a minimum is reached of $\approx 0.94
\cdot 10^{-2}$ GeV$^{-1}$, after which the amplitude increases again. For
$\phi^{\,\,\prime}_s > 80^\circ$, there would be agreement again with
experiment. However, such a large mixing angle, which would imply an almost
pure $\bar{s}s$ substructure for the \ft\ seems to be excluded by the weak
processes discussed in the previous section, as well as by hadronic decays
\cite{RBS01}.

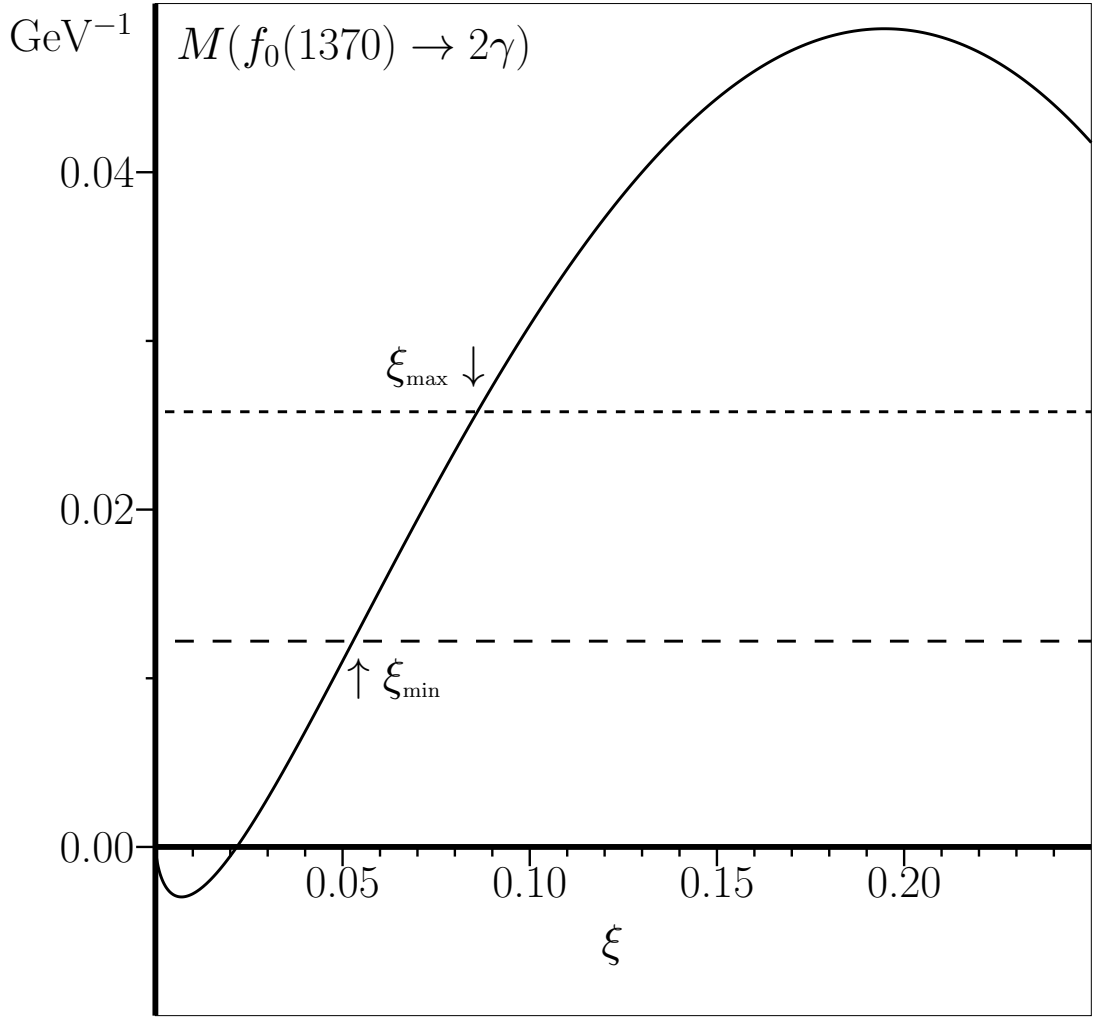
\begin{figure}[hp] 
%\epsfxsize=  17.3cm
%\epsffile{amplitude.eps}
\Large
\begin{center}
\begin{picture}(425.20,435.20)(-72.50,-43.50)
%
%
%
%
\put(70.80,55.85){\makebox(0,0)[tc]{0.05}}
%
\put(141.61,55.85){\makebox(0,0)[tc]{0.10}}
%
\put(212.41,55.85){\makebox(0,0)[tc]{0.15}}
%
\put(283.22,55.85){\makebox(0,0)[tc]{0.20}}
%
%
%
%
%
%
%
%
%
%
%
%
%
%
%
%
%
%
%
%
%
\put(-8.00,63.85){\makebox(0,0)[rc]{0.00}}
%
\put(-8.00,191.56){\makebox(0,0)[rc]{0.02}}
%
\put(-8.00,319.27){\makebox(0,0)[rc]{0.04}}
%
%
\put(172.99,34.75){\makebox(0,0)[tc]{$\xi$}}
\put(-8.00,379.11){\makebox(0,0)[tr]{GeV$^{-1}$}}
\put(8.00,375.10){\makebox(0,0)[tl]{$M(f_0(1370)\rightarrow2\gamma)$ 
\raisebox{-4.25cm}{\hspace*{-2.3cm} $\xi_{\mbox{\scriptsize max}}\downarrow$}
\raisebox{-8.4cm}{\hspace*{-2.2cm} $\uparrow \xi_{\mbox{\scriptsize min}}$}}}
%
%
%
\end{picture}
\end{center}
\normalsize
\caption[]{Two-photon-decay amplitude of the $f_0(1370)$ determined by
$u,d$ \/quark loops.
Here, $\xi =m_{n}^{2}/m_{f_{0}(1370)}^{2}$, with $m_{n}$ representing the
constituent nonstrange quark mass $m_{u}=m_{d}$, and $\xi_{\em min}$, $\xi_{\em
max}$ stand for the one-standard-deviation boundaries of the experimental
estimate given in Eq.~(17) for the two-photon decay rate of the $f_0(1370)$
meson. The corresponding nonstrange quark masses range from 315 to 402 MeV.}
\label{fig1}
\end{figure}

Alternatively, if instead we try the $\bar{n}n$ $\;u,d$ quark loops,
the $f_0(1370)\rightarrow2\,\gamma$ amplitude would be \cite{DEMSB94}, for
$\xi\simeq m^2_u/m^2_{f_0(1370)} \simeq m^2_d/m^2_{f_0(1370)} \le 1/4$,
\begin{eqnarray} M(f_0(1370) \rightarrow 2\,\gamma) & = & \sqrt{2}\;\,
\mbox{Tr} \, [Q^{2\,} \, Q_{\bar{n}n}] \;\; \frac{\alpha \, N_c}{\pi\, F_\pi}
\;\; 2\, \xi \; [\,2+(1-4\,\xi) \, I(\xi)\,] \nonumber \\
 & = & \frac{5\,\alpha \, N_c}{9\, \pi\, F_\pi} \;\; 2\, \xi \; [\,2+(1-4\,\xi)
 \, I(\xi)\,] \, .\label{eq14}
\end{eqnarray}
For $\xi<1/4$, the values $0.053 < \xi \simeq m^2_u/m^2_{f_0(1370)}
\simeq m^2_d/m^2_{f_0(1370)} < 0.086$ are compatible with the experimental
estimate in Eq.~(\ref{eq10}), i.e., $|M(f_0(1370)\rightarrow 2\,\gamma)| =
(1.90 \pm 0.68)\cdot 10^{-2}\;\mbox{GeV}^{-1}$. For $m_{f_0(1370)} \simeq 1370$
MeV, the allowed ranges for $\xi< 1/4$ yield $315 \;\mbox{MeV} < m_u \simeq
m_d < 402 \;\mbox{MeV}$ (see Fig.~\ref{fig1}). Using $I(\xi)$ given in
Eq.~(\ref{eq9}), we observe that for all $\xi>1/4$, which would anyhow imply
unrealistically large quark masses, the quark-loop rate is not consonant with
the experimental estimate. The allowed range for the
constituent $u$,$d$ mass is quite consistent with the $f_0(1370)$ being purely
$\bar{n}n$, or with a small $\bar{s}s$ admixture, of course. On the other hand,
taking the \ft\ to be mostly $\bar{s}s$, it is almost impossible to find any
reasonable quark masses and mixing angles to get agreement with experiment.

\section{Conclusions} \label{sec4}
In this paper we have studied weak and electromagnetic decay processes with
isoscalar scalar mesons in the final and initial state, respectively, in order
to identify the quark substructure of especially the \fn, \ft, and \ff\
resonances.

Calculating the weak process $D_s^+\rightarrow f_0\pi^+$, which
has been observed for the \fn, \ff, and \fs, via the standard $W^+$-emission
graph, leads to good agreement with experiment for the \fn\ and \ff, if
these states are assumed to be mostly $\bar{s}s$. For the \fs, the large
experimental error does not allow a definite conclusion about a possible
dominant $\bar{s}s$ configuration, but a mostly $\bar{n}n$ substructure of this
resonance is unlikely. As to the \ft, the PDG tables do not report the process
$D_s^+\rightarrow \ft\pi^+$ at all, which would exclude a mostly $\bar{s}s$
nature of this resonance. Not even the observation of the process by the E791
collaboration seems to affect this conclusion, since $D_s^+\rightarrow \ft\pi^+
\rightarrow K^+K^-\pi^+$ is \em not \em \/ observed \cite{G00}.

Regarding the electromagnetic processes, calculation of the experimentally
observed two-photon decays $\fn\rightarrow\gamma\gamma$ and $\ft\rightarrow
\gamma\gamma$, using either quark or meson loops, leads to good agreement with
the experimentally measured rates, provided that the \fn\ is assumed to be
mostly $\bar{s}s$ and the \ft\ mainly $\bar{n}n$, and taking moreover the
controversial PDG data on the \ft\ at face value (see discussion in Sect.~3.2).
While the quark-loop results
depend rather sensitively on the (model-dependent) quark masses and mixing
angles, especially in the case of the \fn, the meson-loop results only
depend on the $\bar{n}n$ vs.\ $\bar{s}s$ mixing and, therefore, are more stable
and reliable. 

At this point we should remark that, in a strict SU(3) extension of the
quark-level LSM (qlLSM) \cite{DS98}, which to some extent underlied our
approach here,
both quark \em and \em \/meson loops should be included in the two-photon decay
amplitude of the \fn, being a ground-state scalar meson. As a matter of fact,
the contributions of both kinds of loops are needed for the $\sigma$(600) ---
in the SU(2) case ---  so as to get near the not-so-well known experimental
two-photon width of the \fsigma\ (see Ref.~\cite{S99}, reference no.\ 19).
However, as mentioned in the text, the quark-loop result for the \fn\ is very
sensitive to the quark masses and the mixing angle, due to a rate-enhancement
factor of 25 for the nonstrange $\bar{q}q$ component. By a judicious but not
unreasonable choice of these parameters, one can easily make the quark-loop
contribution vanish, which would occur (using $g_{{f_0\,SS}} = \sqrt{2}\;2\pi
/\sqrt{3}$) for e.g.\ $m_{u,d}=340$ MeV, $m_s=490$
MeV, $\phi_s=12.4^{\circ}$, or $m_{u,d}=300$ MeV, $m_s=432$ MeV,
$\phi_s=18.3^{\circ}$, or all kinds of intermediate values. Therefore, our
conclusion on the dominantly $\bar{s}s$ nature of the \fn\ is upheld no matter
which framework is used, i.e., either the rigorous SU(3) qlLSM or the more
phenomenological meson-loops-only approach.

Summarizing, weak and electromagnetic processes lend quantitative evidence to
a dominantly $\bar{s}s$ interpretation of the \fn\ and \ff, and a mostly
$\bar{n}n$ assignment for the \ft.
\vspace{1cm}

{\it Acknowledgements.} 
We wish to thank D.~V.~Bugg and M.~R.~Pennington for valuable exchanges of
ideas. This work was partly supported by the
{\em Funda\c{c}\~{a}o para a Ci\^{e}ncia e a Tecnologia} (FCT) 
of the {\em Minist\'{e}rio da Ci\^{e}ncia e da Tecnologia} of 
Portugal, under Grant no.\ PRAXIS XXI/\-BPD/\-20186/\-99 and under contract
numbers POCTI/\-35304/\-FIS/\-2000 and CERN/\-P/\-FIS/\-40119/\-2000.

\end{document}